# A SPATIAL PREDICTIVE MODEL FOR MALARIA RESURGENCE IN CENTRAL GREECE INTEGRATING ENTOMOLOGICAL, ENVIRONMENTAL AND SOCIAL DATA

**Panagiotis Pergantas[1], Andreas Tsatsaris[2], Chrisovalantis Malesios[3*], Georgia Kriparakou[1], Nikos Demiris[4], Yiannis Tselentis[5]**

[1]*Bioapplications Ltd.*

[2]*Technological Educational Institute of Athens, Department of Department of Civil, Surveying and Geoinformatics Engineering*

[3*] *(corresponding author) Democritus University of Thrace, Department of Rural Development, e-mail:* [malesios@agro.duth.gr](malesios@agro.duth.gr)

[4]*Athens University of Economics and Business, Department of Statistics*

[5]*University of Crete, Regional Public Health Laboratory, Faculty of Medicine*

**Short title:** A spatial predictive model for malaria resurgence

ABSTRACT

Malaria constitutes an important cause of human mortality. After 2009 Greece experienced a resurgence of malaria. Here, we develop a model-based framework that integrates entomological, geographical, social and environmental evidence in order to guide the mosquito control efforts and apply this framework to data from an entomological survey study conducted in Central Greece. Our results indicate that malaria transmission risk in Greece is potentially substantial. In addition, specific districts such as seaside, lakeside and rice field regions appear to represent potential malaria hotspots in Central Greece. We found that appropriate maps depicting the basic reproduction number, $R_0$, are useful tools for informing policy makers on the risk of malaria resurgence and can serve as a guide to inform recommendations regarding control measures.

# 1 INTRODUCTION

Malaria is one of the well-studied vector-borne diseases in terms of its transmission and the potential for transmission's change. Malaria is endemic in about 100 countries around the world, mainly in sub-Saharan Africa and Asia (*WHO, 2015*). It is a mosquito-borne parasitic infectious disease, transmitted through the bite of the infected female *Anopheles* mosquito. Five types of plasmodia cause disease to humans: *Plasmodium falciparum*, *Plasmodium vivax*, *Plasmodium ovale*, *Plasmodium malariae* and *Plasmodium knowlesi,* of which *P. falciparum* and

*P. vivax* are the most prevalent and *P. falciparum* the most dangerous (*HCDCP, 2012*). In 2015 the number of malaria deaths globally were 438000 whereas 3.2 billion people being at risk, with more deaths occurring in Africa (90%), followed by South-East Asia (7%) and the Eastern Mediterranean region (2%) (*WHO, 2015*). (Climatic determinants are considered particularly important for malaria, since both the disease agent (*Plasmodium*) and vectors (*Anopheles* mosquitoes) are strongly affected by climate (*Craig et al., 2004*). Global warming has been considered as a potential risk for malaria resurgence in northern hemisphere areas (*Lindsay and Thomas, 2001*). Therefore, most studies have used statistical relationships between malaria transmission or vector occurrence and climate in order to project the potential future distribution of malaria transmission areas (*Leedale et al., 2016*).

Epidemics in general, can be caused, except from climate conditions, by many other factors, for instance movement and displacement of human populations. Recently, an increasing number of studies point towards a shift from climate as the main driver for malaria occurrence, to additional non-climate factors as responsible for malaria occurrence and transmission (*Hay et al., 2002; Midega et al., 2012*; *Thomas et al., 2013*). Others suggested a more realistic approach, proposing that this complex system of malaria transmission is based on a combination of climate and non-climate factors (*Hardy et al., 2015*). A notable such factor relates to movement and displacement of human populations. Human migrants infected elsewhere can move, temporarily or permanently, into a new non-infected region or country. There is a significant body of literature that concentrates on mobile populations and their importance as a malaria transmission factor (see, e.g., *Gomes et al., 2016*; *Pindolia et al., 2012*; *Smith and Whittaker, 2014*). Understanding the spatio-temporal distribution of risk for mosquito-borne infections, such as malaria, is important (*Smith et al., 2004*).

Free movement within the European Union has been largely facilitated by an environment essentially free of important diseases. However several European regions have recently seen an influx of new-coming populations, including migrants and refugees. This fact, combined with natural habitats which can serve as hot spots for malaria resurgence due to their suitability in terms of mosquito populations, suggest that increased vigilance is required in order to avoid the (re-)establishment of previously endemic diseases (*Piperaki and Daikos, 2016*). Already in the years 2011 and 2012, in Southeast Europe renewed malaria transmission was observed, with Greece and Turkey counting localized outbreaks as a result of malaria importation from endemic countries (*WHO, 2015*).

Malaria has been eradicated from Greece in 1974 after coordinated efforts by the WHO and the local authorities (*Danis et al., 2013*). Over the last decade a large population of migrants from countries were malaria is endemic came to Greece. As a result, in the recent years, a small number of imported cases of malaria have been reported annually (*HCDCP, 2015*). Although these cases are sporadic, they raise the chance of local transmission (*Kampen et al., 2003*). Specifically, in 2009 and 2010 some cases of *P. vivax* malaria (51 and 44, respectively) were laboratory confirmed in Greece (mainly in the agricultural area of Evrotas, Laconia in Peloponnese in Southern Greece). In 2011, an outbreak reached a total number of 96 laboratory confirmed *P. vivax* cases of which the 54 were imported cases of migrants from malaria endemic countries, and 42 were domestically transmitted cases (*HCDCP, 2015*). In 2012, a total of 93 laboratory confirmed cases of malaria were reported in Greece, of which 73 were imported (64 in migrants from malaria endemic countries). After the 2011 outbreak, a multidisciplinary strategy with a variety of intensive response activities, was adopted and implemented in Evrotas (*Danis et al., 2013*). Despite the fact that during 2014 Greece had no domestically transmitted cases of malaria, there were 6 new cases in 2015, indicating the need for constant vigilance in the region (*WHO, 2015*). See Table 1 below for an analytical description of malaria cases during 2009-2015.

| Year | Imported cases | Seemingly domestically transmitted cases | Plasmodium |
|---|---|---|---|
| 2009 | 44 | 7 | *P.vivax* |
| 2010 | 40 | 4 | *P.vivax* |
| 2011 | 54 | 42 | *P.vivax* |
| 2012 | 73 | 20 | *P.vivax* & *P.falciparum* |
| 2013 | 22 | 3 | *P.vivax* & *P.falciparum* |
| 2014 | 38 | 0 | *P.vivax* |
| 2015 | 79 | 6 | *P.vivax* |

**Table 1.** Total numbers of malaria confirmed cases in Greece (period: 2009-2015).

From the information presented above, we may recognize that malaria represents an important emerging public health issue in Greece, particularly in the Central and Southern regions of mainland Greece (*Danis et al., 2013; Tseroni et al., 2015*). A malaria resurgence risk in Greece varies dramatically with space and depends, among other factors, on environmental changes as well as the introduction of parasite carriers. Therefore, it is imperative that malaria early-warning systems are put in place to enhance public health decision making for prevention and control of malaria epidemics. This aim can be achieved via the development of a prediction model which incorporates all the relevant sources of evidence, including mosquito abundance and potential disease transmitters.

In the current paper, in order to determine the potential transmission risk of malaria resurgence in Greece, we develop and present an appropriate host-vector spatially explicit model which integrates entomological, geographical, social and environmental evidence in order to guide the mosquito control efforts. This model, which is a suitably tailored version of the classical Ross-Macdonald model (*Macdonald, 1952*) combines entomological, environmental and social data. We present spatial maps with three distinct but complementary severity scales; (i) the basic reproduction number, which is a combination of human to mosquito and mosquito to human transmission and represents the (mosquito-driven) disease potential from one (infected) human host to other humans, (ii) the probability of getting infected, which incorporates the initial number of infected humans and (iii) the expected number of infected cases.

## 2 MATERIALS AND METHODS

### 2.1 GIS Database and data collection

*Study Area*

The entomological evidence for this study is informed by a large-scale mosquito abatement program conducted by Bioapplications Ltd during the last three years in central Greece. The study took place in 8 municipalities of the Prefecture of central Greece (see Figure A1 in the Appendix). The study area covers about 406.000 ha.

We established "MALGDB" as a Malaria Geographical Database which gathers together geographical, environmental, population, entomological and epidemiological data. We used ArcGIS 10.3 software to create, develop and populate that database. Nomad rugged handheld computers running Arcpad 7.1 software were used to gather field data and information. All data collected were sent via GPRS connection to the main MALGDB. The database that was build contains administrative units, populated places, digital elevation model, land cover, surface temperature and humidity, population density and composition, # of migrants from malaria endemic countries, larvae and adult *anopheles* positions and population, human cases, etc.

As concerns the species involved in the study, it has been found after morphological examination that the species *Anopheles sacharovi*, *An. maculipennis*, *An. superpictus, An. claviger* and *An. hyrcanus* are the most important species in the study region (account for approximately 90% of all *Anopheles* species in the region). It should be noted that all these types of identified species are potential malaria vectors. According to the above, in our study we suppose that all the positive samplings for *Anopheles* mosquito larvae are larvae belonging to species that can transmit malaria.

As regards the larval sampling, net mesh has been utilized. The diameter of the net mesh was 20 cm and for each sample we scanned 5 different lines of 1 meter long each. Based on the morphology of the natural (relief, streams etc.) and the human environment (lakes, rice fields, irrigation system, settlements) we selected 10 areas where we had established sample-collecting stations. The sampling stations were situated in the outskirts of towns and villages. Each station consisted of $CO_2$ mosquito traps. Breeding sites included canals, rice pads, tanks etc. Mosquito breeding sites were monitored on a weekly basis from the end of April to the end of autumn in order to record the larval dynamics and some of the physicochemical parameters of breeding sites, such as pH temperature conductivity. In total, there were more than 4000 different mosquito breeding sites in the study areas. Up to 200 different sites were positive at least once for *Anopheles* larvae.

Following collection, mosquito samples were transported to the laboratory, where they were identified and separated by species using morphological identification. The samples provided evidence regarding the variation of the population of *Anopheles* mosquitoes between the different seasons of the year and highlighted the areas where *Anopheles* occurred in high density.

The main malaria reservoirs in Greece relates to migrants from countries where malaria is endemic. Thus, the exact locations of these migrants' habitats were marked up to a GIS database by using field computers. We also kept records of their population in each site. This migrant monitoring was repeated in 3 different periods (in June, early August and mid-September respectively) in order to obtain as possible as accurate data.

The larvae sampling and monitoring was conducted by using digital sampling protocols. We created a user-friendly database (domains in geodatabases), for data collected from the field and loaded in field rugged computers (see Figure 1a, b). In total, the larvae data from the study area comprised of more than 4000 records.

---FIGURE 1 AROUND HERE ---

Every 10 to 15 days we were placing $CO_2$ traps in 10 places. The results from the adult trapping were embodied in the MALGDB. Temperature data were recorded from the meteorological stations of National Observatory of Athens (NOA) (see Figure 2a, b for examples of temperature collection data at varying time periods). There were 10 meteorological stations of the NOA recording temperature data at the study area. For our modeling, we utilized the average temperature values with a 30-day interval, starting from the first Saturday of each month, between the April and November of each year (2012 & 2013). The inverse distance weighting (IDW) interpolation method was additionally used to estimate the average temperatures at the locations where no measurements were available.

---FIGURE 2 AROUND HERE ---

**2.2 Methods**

The main disease severity measure utilized in this paper is the basic reproduction number, $R_0$, which can be interpreted as the expected number of (human) hosts that would be infected by the introduction of an infected host into a large fully susceptible population. The prospects for the success of disease control depend crucially on this measure (*Smith et al., 2007*). Hence, $R_0$ appears to be an appropriate measure and our control strategies are mostly concerned with proportionally reducing $R_0$. We could also calculate the vectorial capacity, a purely entomological equivalent of $R_0$, defined as the expected number of mosquito bites that would eventually arise from all the mosquitoes which would bite an infectious individual on a single day. However, since the mobile populations appear to be an important potential driver of transmission we mostly focus upon $R_0$ since it incorporates the vectorial capacity as well as the exposure due to potentially infected human hosts within a single measure. Some of the parameters necessary for $R_0$ estimation were simulated based upon values reported in the literature and these results were fused with entomological information collected in the field by Bioapplications Staff (see Table A1 in the Appendix for the parameter details).

**2.3 The spatial predictive model**

In this paper we utilize a host-vector model that combines entomological, environmental and geographical data to provide estimates on the average infection number due to malaria in central Greece. To achieve this, the methodology proposed here takes into account the potential host population in the region, related to migrants from countries where malaria is endemic, in addition to the standard entomological parameters which are based on the well-established Ross-Macdonald mathematical model of mosquito-borne pathogen transmission (*Smith and McKenzie,*

*2004*).

Specifically, since the number of expected infections for the different regions ($i = 1, 2, ..., 710$ and $i = 1, 2, ..., 183$ for each of the years 2012 and 2013, respectively), say $E(\text{infections})$ is given by:

$$E(\text{infections}) = \Pr(\text{infection}) \times (\# \text{ of susceptibles}),$$

we seek to estimate the probability of infection, $\tau = \Pr(\text{infection})$. If we denote with $\hat{\tau}$ this estimate, let us assume that the latter is a function of two different measures, one describing the disease potential due to mosquito abundance and the other the component attributed to host infections. The reproduction number $R_0$ is utilized for the former whereas for the latter transmission route we use the proportion of initially infected human hosts, denoted by $\mu_0$.

For the calculations concerning $\hat{R}_0$, which describes the expected number of hosts that would be infected by a single infected host in a large susceptible population, we are obtaining a different estimate of $R_0$ for each region, calculated as:

$$\hat{R}_{0i} = \frac{V_i \cdot b_i \cdot c}{r_i}, \tag{1}$$

where $V$ denotes the vectorial capacity, i.e. the expected number of infective mosquito bites that would eventually arise from all the mosquitoes that would bite a single fully infectious person on a single day (*Smith et al., 2012*), and is given by:

$$V_i = \frac{m_i \cdot \alpha_i^2 \cdot \exp(-g_i \cdot v_i)}{g_i}. \tag{2}$$

In equations (1) and (2), $m_i$ denotes the ratio of mosquitoes to humans in each region $i$; $\alpha_i$ the biting rate, i.e. the proportion of mosquitoes that feed on humans each day; $b_i$ the probability a bite produces infection to a human; $r_i$ the average recovery rate per day; $v_i$ the mosquito latent period, i.e. the number of days from infection to infectiousness; $g_i$ the mosquito instantaneous death rate per day. Finally, with $c$ we denote the probability a mosquito becomes infected after biting an infected human, which for our analysis is set to the constant value 0.5.

The parameters $\alpha_i$, $b_i$, $r_i$ and $v_i$ were sampled from suitable distributions according to the relevant literature (see Table A1 in the Appendix for a detailed description of model parameters and corresponding values and distributions assigned).

For the purposes of the current study, we modify equation (2) to account for the more realistic assumptions of temperature-dependence of the mosquito latent period ($v_i$) and the mosquito instantaneous death rate per day ($g_i$). Hence, since $v_i$ is known to be strongly dependent on temperature, $T_i$, (see *Paaijmans et al., 2009*; *Paaijmans et al., 2013*; *Mordecai et al., 2013*), we incorporate the latter dependence into our modeling via the following equation:

$$V_i = \frac{m_i \cdot \alpha_i^2 \cdot \exp\left[-g_i \left(\delta_i + v_i + \alpha_i^{-1}\right)\right]}{g_i}, . \tag{3}$$

where the pre-blood meal period (or infection delay), $\delta_i$, is a function of temperature through:

$$\delta_i = 0.0163 \cdot T_i^2 - 0.95 \cdot T_i + 14.769.$$

The above relationship has been obtained through experiments on *Anopheles* mosquitoes (*Paaijmans et al., 2013*). Note also that in order to calculate $g_i$ we are assuming that it changes with the temperature levels (*Smith et al., 2004*), since that temperature is known to influence the mosquito life cycle and in particular the development rate of larvae and adult survival (*Bayoh and Lindsay, 2003*), hence:

$$g_i = \frac{1}{-4.4 + 1.31 \cdot T_i - 0.003 \cdot T_i^2},$$

where each $g_i$ corresponds to the maximum of the calculated values for each one of the different temperatures, i.e. $g_i = \max\{g_{ij}\}$, for temperatures $T_i$ recorded at $i = 1, 2, .., 30$ [see also *Brady et al. (2016)* for a study on how to modify the mosquito daily mortality rate $g_i$ to account for the immature mosquito stages].

As regards the estimation of the external host component due to the migration, denoted by $\mu_0$, expressing the proportion of initially infected hosts in each one of the 710 and 183 sub-regions for 2012 and 2013 respectively, we use the following approximation:

$$\hat{\mu}_{0i} = \sum_{k=1}^{3} \mu_{0ik} \cdot W_{ik},$$

where $W_{ik}$ is a simple exponential kernel function of the form:

$$W_{ik} = alpha \cdot \exp(-alpha \cdot d_{ik}),$$

that is used to model the spatial component of migrant transmission, through the distances $d_{ik}$ from the larvae areas, measured during the three periods of migrant monitoring $k = 1,2,3$. The prevalence of asymptomatic malaria varies widely between and within country (see e.g. *Imwong et al. 2015*; *Starzengruber et al., 2014*; *Waltmann et al., 2015*) and also depends upon the detection method. Therefore, we used three distinct scenarios where the baseline prevalence was set to 10% and we also examined a prevalence of 5% and 20% in deterministic sensitivity analyses. Hence, the estimated proportion of initially infected hosts, $\hat{\mu}_{0i}$, is multiplied with the latter pre-specified incidence rates.

Having obtained the $\hat{R}_{0i}$ and $\hat{\mu}_{0i}$ estimates for each region, we proceed to the estimation of $\hat{\tau}$ by solving the following non-linear equation for $\hat{R}_{0i} \geq 1$:

$$1 + \hat{\mu}_{0i} - \hat{\tau}_i - \exp(-\hat{\tau}_i \cdot \hat{R}_{0i}) = 0,$$

whereas for $\hat{R}_{0i} < 1$ we set $\hat{\tau}_i = 0$. Accordingly, we can estimate the $E(\text{infections})$. We have used the R package (*R Development Core Team, 2008*) for data manipulation and in order to fit the above described model to central Greece malaria resurgence data for the years 2012 & 2013. The R code of the spatio-temporal model is available upon request by the corresponding author. The data for the years 2012 & 2013 can be obtained from S1 and S2 supplement files, respectively.

In order to aid the illustration of our results, we construct $R_0$ maps that can be used for informing policy makers on the risks of malaria resurgence. Similar maps, in the case of other infectious viruses, have been presented elsewhere; see for instance *Hartemink et al. (2009)* for an application of $R_0$-based maps on the bluetongue virus.

Under the modeling framework described in the previous section we calculated estimates of the basic reproductive number ($R_0$) for each of the regions of Central Greece (710 and 183 for the years 2012 and 2013, respectively), for the periods between April and November. The basic reproductive number provides a threshold criterion. When $R_0 >1$ the disease may spread over a large part of the population while when $R_0 <1$ the disease may only affect a small proportion of the population. The estimated $R_0$ values showed marked differences between the regions of central Greece, in both years. The density surfaces in the $R_0$ geographical maps are based upon point estimates. In particular, the maps shown in figures 3 and 4 present the median $R_0$ values above the threshold of 1 across the two years.

An altitude zone that includes the areas between 0 and 300 m was applied as a threshold in order to avoid the depiction of the density in areas which are unlikely to be affected from the represented phenomenon. Therefore, we decided to effectively preclude interpolation methods since those may or may not be appropriate in our setting due to the presence of intermediate areas of high altitude which could have been depicted as high risk when, in fact, this is not the case.

The method used to map the results was "Kernel Density" (Spatial Analyst of ArcGIS – ArcMap v.10.3). The Density tool distributes a measured quantity of an input point layer throughout the landscape to produce a continuous surface. These geographical maps can be used to identify the areas of higher risk for a malaria outbreak after the introduction (see a detailed explanation of the method in S3 supplement file). As we observe, for both years there exists a geographical similarity as regards the potential malaria transmission hotspots.

**3 RESULTS**

Approximately ten distinct geographical units in malaria resurgence can be identified by inspecting the two graphs. Specifically, the areas of higher resurgence risk can be identified around the cities of Lamia, Levadeia and Chalkida, as well as in the Northern and Southern parts of Evoia. Generally, risk of malaria resurgence is increased in seaside areas, areas near lakes where the *Anopheles* mosquito population density peaks (e.g. Chalkida), or at the *Anopheles* breeding habitats such as areas with paddies (e.g. Lamia).

---FIGURE 3 AROUND HERE ---

The findings point to the fact that resurgence is possible in both areas of sparse human populations (rural areas) and areas with dense human populations (urban areas). In case of disease resurgence or outbreak, interventions could be targeted in these areas identified by our analysis as regions of high transmission intensity.

--- FIGURE 4 AROUND HERE ---

Another important − complementary to the $R_0$ − factor in disease modeling is the probability of infection, $\tau$, and accordingly the amount of people that are expected to be infected in case of the disease resurgence in the areas under investigation. These two outcomes are complementary but add extra source of information, hence we additionally present the results of the two outcomes in the Appendix. Figures A2 to A5 show the latter measures, for the years 2012 and 2013, respectively. We can see from the inspection of the $\hat{\tau}$ estimates (Figures A2 &

A3) that the human populations with the higher risk for infection are located mainly in the wider region of the city of Lamia, and in the region of Northern Evoia. This is a result that seems to be time invariant for both years of research. In addition, a high risk for infection is also present in the seaside areas of the Fthiotida region. Higher values of $\hat{\tau}$ are also found in southern Evoia for the 2013 year.

Next, as regards the model estimates of the numbers of expected infections from malaria (see Figures A4 & A5), we observe similar results between the years 2012 and 2013. The highest amounts of people expected to be infected are widely scattered in the regions of central Greece, with higher at the city of Lamia and neighboring regions. We also observe high values for northern Evoia and other seaside areas.

As expected, malaria transmission is highly seasonal, with transmission limited to the warm and rainy summer months. Cases in general increase by the end of June, peak in late summer and decline by the end of November. The average seasonal pattern in malaria incidence follows the periodicity in rainfall and temperature, with a 3 to 4 month lag (*Craig et al., 2004*).

In the following table (Table 2) the summary values of median $R_0$ are presented, for various periods of the two years as were estimated from our model, thus giving a picture of the seasonality pattern. Besides the total median $R_0$ estimates, we also report the median $R_0$ for both the urban and rural regions of Central Greece. As we observe, there is a seasonal trend of $R_0$ increase in both years, occurring during the summer months when the temperatures are obviously higher. The $R_0$ values are showing a decline by October-November, where the transmission potential is lower. We note that the major peaks in potential risk resurgence are in July, August and September.

| Year | | Apr | May | Jun | Jul | Aug | Sep | Oct | Nov |
|---|---|---|---|---|---|---|---|---|---|
| 2012 | Total | 0.078 | 0.266 | 0.585 | 0.906 | 1.058 | 1.068 | 0.740 | 0.011 |
| 2012 | Urban | 0.043 | 0.138 | 0.154 | 0.365 | 0.423 | 0.422 | 0.288 | 0.039 |
| 2012 | Rural | 0.136 | 0.334 | 0.801 | 1.255 | 1.449 | 1.437 | 1.011 | 0.017 |
| 2013 | Total | 0.808 | 0.715 | 1.140 | 1.253 | 1.231 | 1.270 | 0.143 | --- |
| 2013 | Urban | 0.257 | 0.333 | 0.393 | 0.396 | 0.295 | 0.396 | 0.072 | --- |
| 2013 | Rural | 0.851 | 1.028 | 1.353 | 1.450 | 1.456 | 1.430 | 0.226 | --- |

**Table 2.** Median basic reproductive numbers by month/region type (Years 2012 & 2013).

It is also noteworthy that malaria resurgence is more likely to occur in rural regions of Central Greece. Tables A2 & A3 in the Appendix present the corresponding estimates for the probabilities of infection $\tau$ and the average number of potential infections in the human populations in case of the disease resurgence. It is found that for the year 2012, the average number of potential human infections ranges between 15 and 17 for the complete populations, whereas the numbers are slightly increased in the case of rural regions of central Greece. The highest values are shown during the period between May and October. The results for the year 2013 are elevated, with the favorable transmission period being between June and September, with approximately 24 and 26 human infections, for the total population and rural regions, respectively.

Finally, Table A4 in the Appendix shows the results of sensitivity analysis where we are varying the value of the baseline prevalence. As seen in Table A4, the three outcomes do not alter significantly from the changes in the baseline prevalence.

## 4 DISCUSSION AND CONCLUSIONS

We adopted a model-based framework that integrates entomological, geographical, social and environmental evidence in order to examine the potential for malaria resurgence in Greece. In addition, disease risk maps were generated in order to assist the interpretation of the results.

The results clearly indicate a potential malaria transmission risk in Greece. Our results characterize the higher and lower risk areas of malaria resurgence in parts of central Greece. We term as hotspots for malaria resurgence those areas with $R_0 > 1$. It has been shown that there are spatial variations in the risk for malaria resurgence. Similar results have been reported in another recent study modeling the risk of malaria resurgence using a mathematical model in Portugal (*Gomes et al., 2016*). Malaria resurgence in Greece may occur, even in areas that are currently free from the disease. Among the main findings is that at elevated temperatures the new-emergence regions are more susceptible to potential outbreaks. Therefore, authorities should be vigilant in order to avoid future outbreaks in areas of high disease potential. We also found that the potential for malaria transmission resurgence, especially in the regions identified by our study, is affected not only by the potential of the virus or the specific climate conditions but also by the number of potential disease hosts who are the main drivers of disease risk. Our model highlighted areas of central Greece as being particularly suitable for disease resurgence.

One important question relates to the stability of potential regions for disease resurgence. Our study is indicative of a positive answer to this question, since that the comparative analysis between the two years under study did not showed significant variations. The spatial clusters of high probability for disease resurgence generally do not seem to vary significantly over time. Further research is warranted in this direction since a large time-horizon will give stronger evidence regarding this issue.

The results emphasize the importance of maintaining population and health systems awareness on the potential resurgence of malaria in Greece. Enhanced prevention and control strategies should be planned for rapid implementation in the case of future pathogen importation to prevent malaria resurgence. An approach of this kind ought to be used in order to plan, control, assess and manage a malaria prevention and control program in Greece. In particular, this study suggests a prediction mechanism which should be put in place so that mosquito control programs be efficient in the case of malaria resurgence along with rapid public health response to treat any infection. These two types of action operate in a complementary manner. Therefore, continuous data update and new data monitoring is critical in order to estimate the relative merit of each and assist public health authorities to respond to potential risks of malaria in a cost-effective manner.

One immediately apparent outcome of this work relates to the need to strengthen surveillance and perform integrated mosquito control programs that will help to eliminate the potential risk of malaria reintroduction and reestablishment. We used a contemporary technique which offers control over a massive amount of field data in real time. Consequently, we maximize the effectiveness of the abatement programs in terms of public health risk. This approach suggests a mechanism for efficient and reliable mosquito control programs: use GIS technology and telematics in order to plan, control, assess and modify abatement applications in-situ and in a real-time manner.

Our study has a number of drawbacks. The selected regions were based upon field studies and were not selected within a large randomized experiment. Therefore these results are representative of the studied (and similar) region but further studies are required in order to generalize these findings to wider regions. Also, as with all such field studies, we obtained approximate larvae numbers and a rough estimate of their survival based upon its relationship with temperature. In addition, weekly -- and not daily -- samples were collected so these numbers are subject to sampling bias. Future studies will mark selected mosquitoes and use capture-recapture methodology for more accurate vector estimates. Also, while validation against an external/independent source of evidence is desirable whenever an analysis with potential for policy action is performed, this seemed nearly impossible in our case since we are essentially looking at counterfactual scenarios because what actually happened is the result of different interventions of distinct nature for each area. However, when interest focuses on highlighting the disease potential and the lack of evidence regarding several important aspects of the disease, this can be thought of an acceptable drawback. Finally, we had limited data on the mobility patterns of potential hosts. Further evidence regarding their movement habits would inform a more realistic dynamic movement network leading to more powerful results. Such information, if available, can easily be fused within our model-based framework.

The current study represents an attempt towards an integrated method for predicting risk resurgence of malaria, based on a spatio-temporal mathematical modeling approach, combined with a proposed framework for constructing $R_0$ maps. Future work should include larger geographical areas, both at national and international level, leading to more widely applicable conclusions.

Hay SI, Rogers DJ, Randolph SE, Stern DI, Cox J, Shanks GD, Snow RW (2002) Hot topic or hot air? Climate change and malaria resurgence in East African highlands. Trends in Parasitology 18: 530-534.

Hartemink NA, Purse BV, Meiswinkel R, Brown HE, de Koeijer, A, et al (2009) Mapping the basic reproduction number (R0) for vector-borne diseases: A case study on bluetongue virus. Epidemics 1(3): 153–161.

Hellenic Center for Disease Control and Prevention (HCDCP) (2012) Department of Epidemiological Surveillance and Intervention. Epidemiological data for malaria in Greece, 2012. Available: http://www.keelpno.gr/Portals/0/Files/English%20files/Malaria%20reports/Malaria%20Report_2012_FINAL_23-82013_EN.pdf. Accessed 15 June 2015.

Hellenic Center for Disease Control and Prevention (HCDCP) (2015) Department of Epidemiological Surveillance and Intervention. Epidemiological data for malaria in Greece, 2005 to 2009.

Imwong M, Nguyen TN, Tripura R, Peto TJ, Lee SJ, Lwin KM, et al (2015) The epidemiology of subclinical malaria infections in South-East Asia: findings from cross-sectional surveys in Thailand-Myanmar border areas, Cambodia, and Vietnam. Malaria journal 14:381.doi: 10.1186/s12936-015-0906-x pmid:26424000; PubMed Central PMCID: PMC4590703.

Kampen H, Proft J, Etti S, Maltezos E, Pagonaki M, Maier WA, et al (2003) Individual cases of autochthonous malaria in Evros Province, northern Greece: entomological aspects. Parasitology Research 89: 252–258. doi: 10.1007/s00436-001-0530-2.

Leedale J, Tompkins AM, Caminade C, Jones AE, Nikulin G, Morse AP (2016) Projecting
22

**FIGURES**

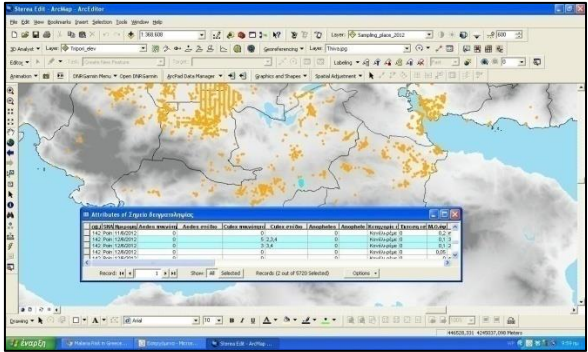 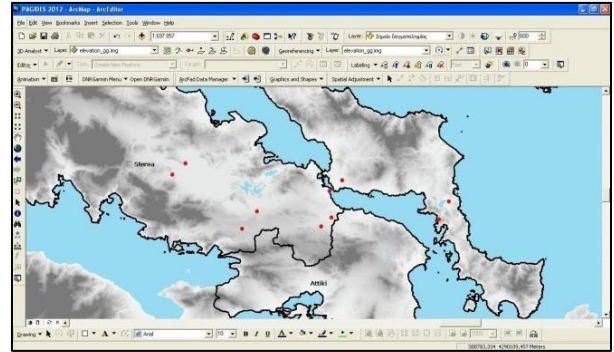

(a) Larvae mosquito sampling stations  (b) Adult mosquito sampling stations

**Figure 1:** Indicative examples of the mosquito sampling stations used in the study (yellow and red dots represent parts of locations for larvae and adult mosquito sampling stations, respectively).

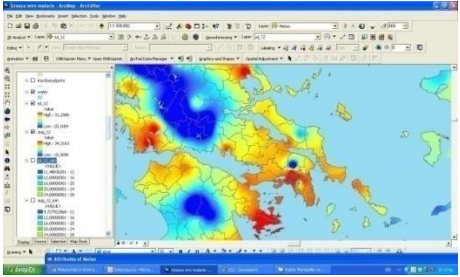 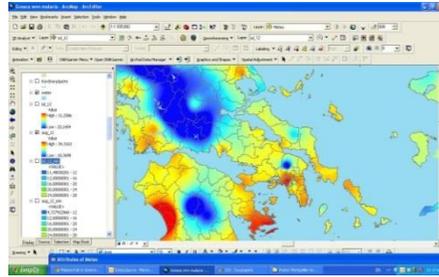

(a) Average temperatures (July 2012)   (b) Average temperatures (August 2012)

**Figure 2:** Indicative examples of calculation and utilization of average temperature data with IDW interpolation method.

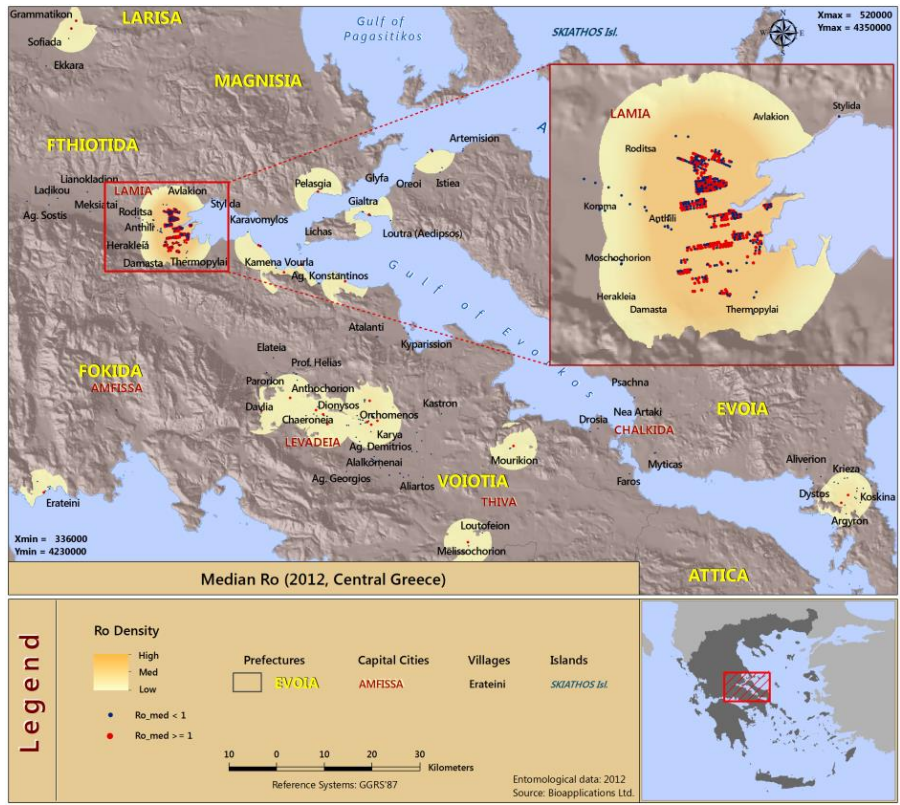

**Figure 3:** Spatial distribution of risk in malaria resurgence in Central Greece as indicated by the median $R_0$ estimates (Year 2012)

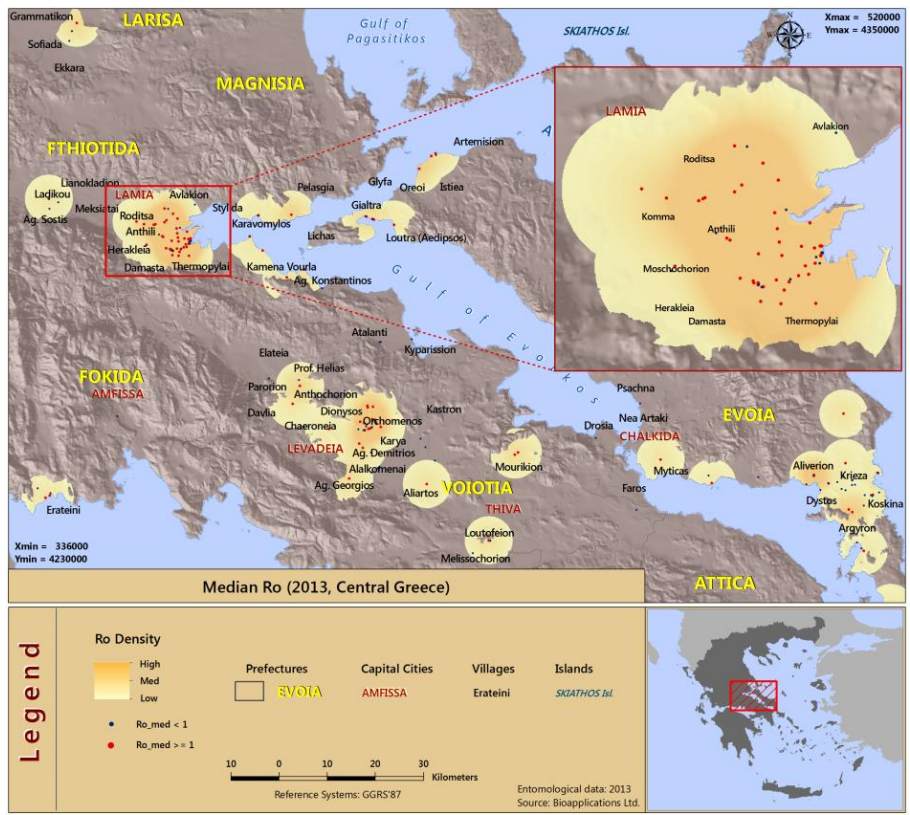

**Figure 4:** Spatial distribution of risk in malaria resurgence in Central Greece as indicated by the median $R_0$ estimates (Year 2013)